\documentclass[11pt,dvips]{article}
\usepackage{epsfig,times} 
\usepackage{picinpar}
\usepackage{wrapfig}
\usepackage{floatflt}
\makeatletter
\renewcommand{\section}{\@startsection{section}{1}{0in}
        {0.4\baselineskip}{0.1\baselineskip}{\Large\bf}}
\renewcommand{\subsection}{\@startsection{subsection}{2}{0in}
        {0.25\baselineskip}{-\baselineskip}{\large\bf}}
\renewcommand{\subsubsection}{\@startsection{subsubsection}{3}{0in}
        {0.1\baselineskip}{-\baselineskip}{\normalsize\bf}}
\makeatother
\newcommand{\app}[3]{#2, Astropart.\ Phys.\ {\bf #1}, #3}

\newcommand{\plb}[3]{#2, Phys.\ Lett.\ {\bf B#1}, #3}

\newcommand{\cpc}[3]{#2, Comm.\ Phys.\ Comm.\ {\bf #1}, #3}
\newcommand{\apj}[3]{#2, Astrophys.\ J.\ {\bf #1}, #3}
\newcommand{\prl}[3]{#2, Phys.\ Rev.\ Lett. {\bf #1}, #3}
\newcommand{\prd}[3]{#2, Phys.\ Rev.\ {\bf D#1}, #3}

\newcommand{\href}[2]{#1}

\begin{document}

\setlength{\baselineskip}{0.95\baselineskip}

%
\thispagestyle{myheadings}
%
\markright{HE 5.1.04}
\begin{center}
%
{\LARGE \bf Indirect Detection of Dark Matter in km-size Neutrino Telescopes}
\end{center}

\begin{center}
%
%
{\bf L.~Bergstr{\"o}m$^{1}$, \underline{J.~Edsj{\"o}}$^{1*}$, 
and P.~Gondolo$^{2}$}\\
{\it $^{1}$Department of Physics, Stockholm University, Box 6730, 
SE-113 85 Stockholm, Sweden\\
$^{2}$Max Planck Institut f\"ur Physik, F\"ohringer Ring 6, 80805
Munich, 
Germany\\
$^*$ presenter}
\end{center}

\begin{center}
{\large \bf Abstract\\}
\end{center}
\vspace{-0.5ex}
%
%
Neutrino telescopes of kilometer size are currently being planned.  
They will be two or three orders of magnitude larger than presently 
operating detectors, but they will have a much higher muon energy 
threshold.  We discuss the trade-off between area and energy threshold 
for indirect detection of neutralino dark matter captured in the Sun 
and in the Earth and annihilating into high energy neutrinos.  We also 
study the effect of a higher threshold on the complementarity of 
different searches for supersymmetric dark matter.
%

\vspace{1ex}

%
%
\section{Introduction}
\label{intro.sec}

Neutrino astrophysics will soon enter a new experimental era.  With
the demonstration by the {\sc Amanda} collaboration (see e.g.\
Halzen, 1999) of the possibility to instrument and successfully
deploy kilometer-long strings with optical modules in the ice cap at
the South Pole station, the road to a km$^3$ detector lies open.  At
the same time, endeavours are underway ({\sc Nestor}, {\sc Antares},
{\sc Baikal}) to prove the possibility of also deploying a large
neutrino detector in a deep lake or ocean.

This new generation of neutrino telescopes will have a larger
effective area than earlier detectors, but the energy threshold will
be higher. A typical energy threshold for these larger detectors is of
the order of 25--100 GeV, and we will here consider thresholds of 1,
10, 25, 50 and 100 GeV\@. As we will show, for the dark matter
detection capability, a low threshold is an important design criterion
to be kept in mind when planning new neutrino telescopes.

As a WIMP candidate we will use the neutralino, that naturally arises
in supersymmetric extensions of the standard model (see e.g.\ Jungman
et al., 1997).

\section{Set of supersymmetric models}

We work in the minimal supersymmetric standard model with seven
phenomenological parameters and have generated about $10^5$ models by
scanning this parameter space (for details, see
Bergstr{\"o}m et al., 1998). 
For each generated model, we check if it is excluded by 
recent accelerator constraints of which the most important ones are the LEP 
bounds (Carr, 1998) on the lightest chargino mass (about 85--91 GeV),
and the lightest Higgs boson mass $m_{H_{2}^{0}}$ (which range from 
72.2--88.0 GeV) and the constraints from $b \to s \gamma$ 
(Ammar et al., 1993 and Alam et al.\ 1995).

For each model allowed by current accelerator constraint we calculate
the relic density of neutralinos $\Omega_\chi h^2$ where the relic
density calculation is done as described in Edsj\"o and Gondolo
(1997), i.e.\ including so called coannihilations. We will only be
interested in models where neutralinos can be a major part of the dark
matter in the Universe, so we restrict ourselves to relic densities in
the range $0.025 < \Omega_\chi h^2 < 0.5$.

\section{Muon fluxes from neutralino annihilations}

The prediction of muon rates is quite involved: we compute neutralino
capture rates in the Sun and the Earth (using the convenient
approximations in Jungman et al.\ (1997)), branching ratios for
different annihilation channels, fragmentation functions in basic
annihilation processes, interactions of the annihilation products with
the surrounding medium (where appropriate), propagation through the
solar or terrestrial medium, charged current cross sections and muon
propagation in the rock, ice or water surrounding the detector.

We simulate the hadronization and/or decay of the annihilation
products, the neutrino interactions on the way out of the Sun and the
neutrino interactions close to or in the detector with {\sc Pythia}
6.115 (Sj{\"o}strand, 1994).  
We treat the interactions of the heavy
hadrons in the centre of the Sun and the Earth in an approximate
manner as given in Edsj{\"o} (1997)\@.

\subsection{Backgrounds and signal extraction.}

The most severe background is the atmospheric background produced by
cosmic rays hitting the Earth's atmosphere (see e.g.\ 
Honda et al., 1995).
For the Sun, there is also a background from cosmic ray interactions
in the Sun (Seckel et al., 1991 and Ingelman \& Thunman, 1996) which
is small but irreducible (at least as long as energy is not measured).

To investigate the possible limits that can be obtained, we will
follow the analysis of Bergstr{\"o}m et al.\ (1997) and parameterize
the neutrino-induced muon flux by
\begin{equation} \label{eq:param}
     \frac{d^2 \phi_s}{dE d\theta} (E,\theta) = \phi_s^0
     \left[ a f_{\rm hard}(m_\chi,E,\theta)+(1-a) 
     f_{\rm soft}(m_\chi,E,\theta) \right] ,
\end{equation}
where $a$ is a model-dependent parameter describing the `hardness' of
the neutrino-induced muon spectrum, $f_{\rm hard}$ and $f_{\rm soft}$
are generic hard and soft muon spectra and $\phi_{s}^{0}$ is the
normalization of the flux.  A typical hard spectrum is given by the
annihilation channel $W^{+}W^{-}$ and a typical soft spectrum is given
by the annihilation channel $b \bar{b}$.  We now assume that the
annihilation spectrum is hard and that $\phi_0$ is the only
unknown. If we relax these assumptions the limits will be up to a
factor of 2--3 higher.

For very high exposures (${\cal E} > 10$ km$^{2}$ yr) towards the Sun,
the above limits will be too optimistic due to the background from
cosmic ray interactions in the Sun's corona. This background will have
about the same angular distribution as the neutralino signal from the
Sun, but quite different energy distribution.  With a neutrino
telescope without energy resolution, this background will put a lower
limit on how small fluxes we can probe from the Sun.  The background
fluxes are about 20, 13, 11, 8.6 and 6.6 muons km$^{-2}$ yr$^{-1}$ for
muon energy thresholds of 1, 10, 25, 50 and 100 GeV respectively. 

\subsection{Dependence on energy threshold.}

\begin{figure}[!t]
\centerline{\epsfig{file=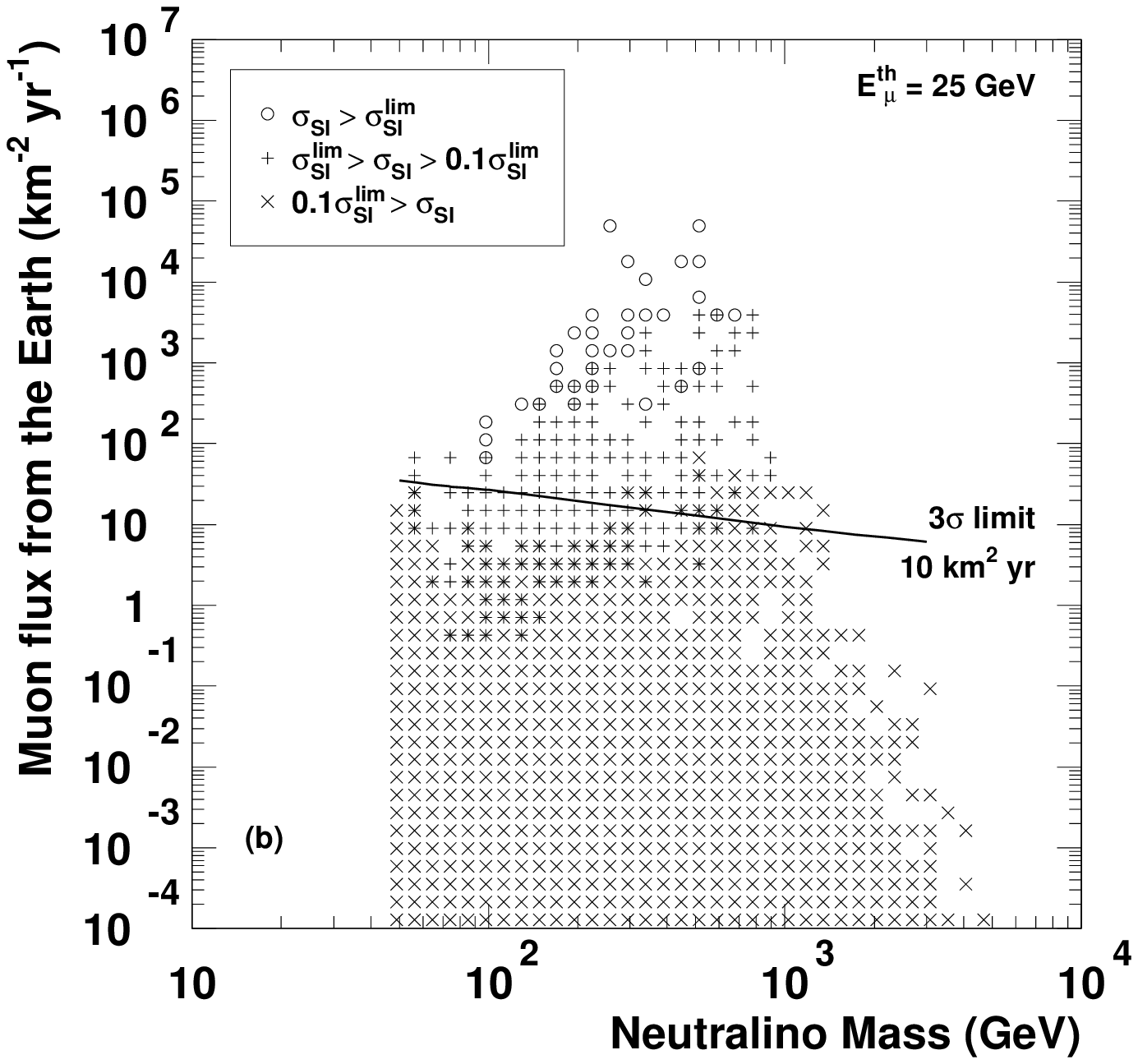,width=0.49\textwidth}
\epsfig{file=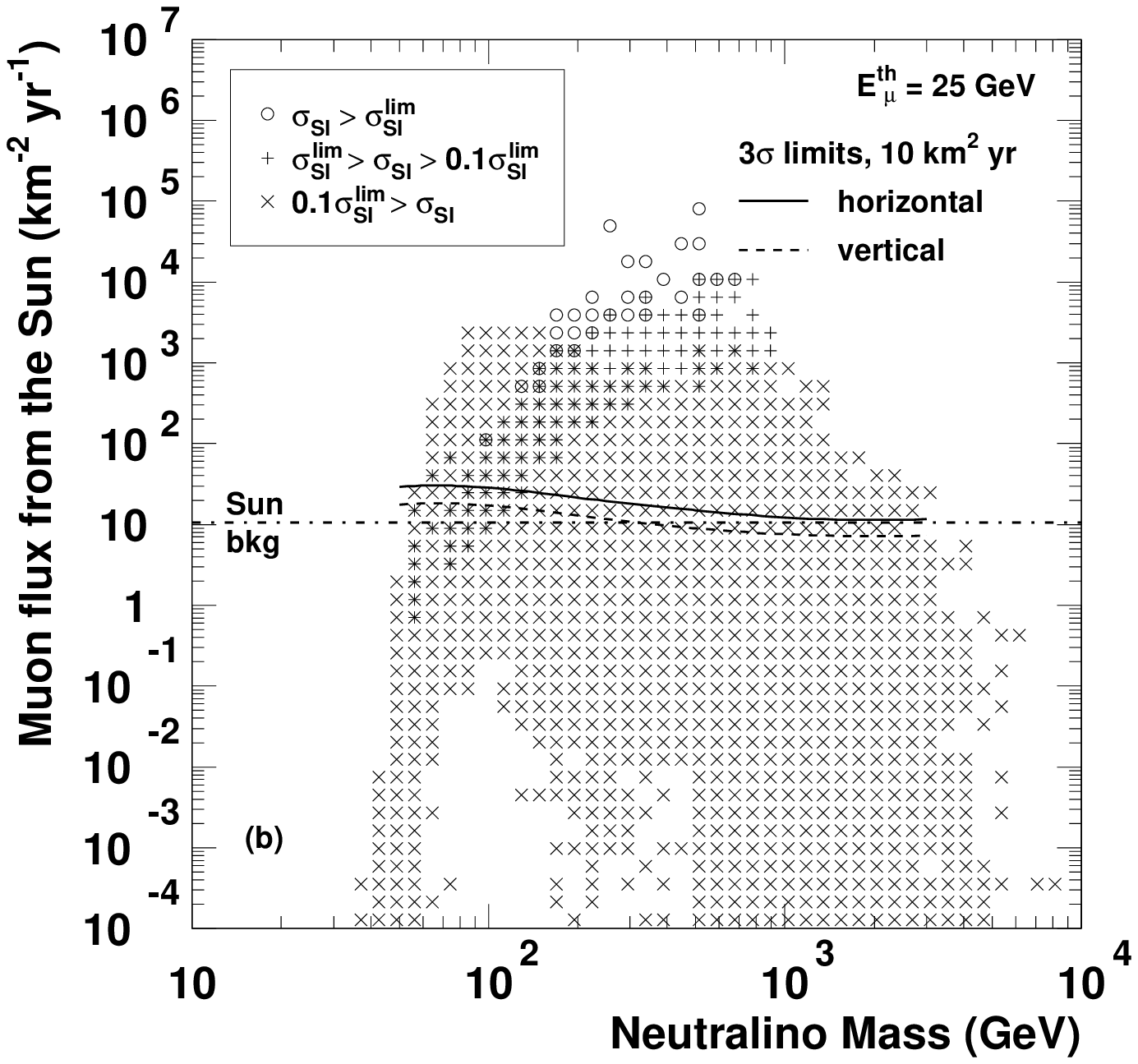,width=0.49\textwidth}}
\caption{The neutrino-induced muon flux from neutralino annihilations
in a) the Earth and b) the Sun. The expected limits that can be
obtained with an exposure of 10 km$^2$ yr are also shown. The models
that can be probed by present direct dark matter searches (Bernabei et
al., 1996), those that can be probed with a factor of 10 increased
sensitivity and those that cannot be probed with direct searches 
are shown with different symbols.}
\label{fig:r25mx}
\end{figure}

As an example of a probably realistic threshold of a km-size neutrino
telescope, we choose 25 GeV and in Fig.~\ref{fig:r25mx} (a) and (b) we
show the muon fluxes versus the neutralino mass. We also show the best
limits obtainable with an exposure of 10 km$^{2}$ yr, and for the Sun,
the background from cosmic ray interactions in the Sun's corona. Note
that for high masses, there is no need to go above an exposure of
about 10 km$^2$ yr towards the Sun (unless the detector has good
energy resolution) due to the irreducible background from the Sun's
corona. For lower masses, the corresponding exposure would be $10^2$
km$^2$ yr.

\begin{figure}
\centerline{\epsfig{file=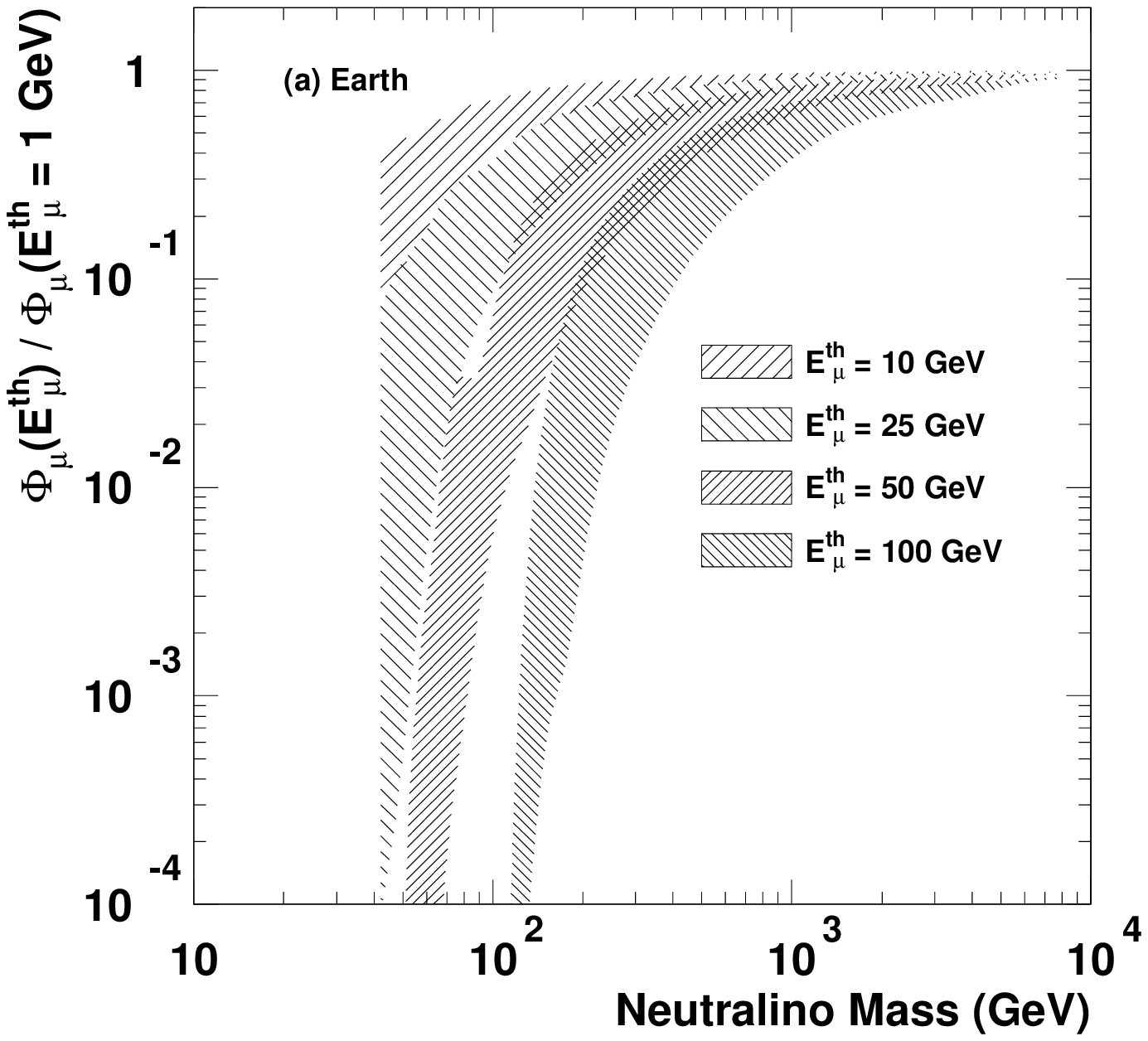,width=0.49\textwidth}
\epsfig{file=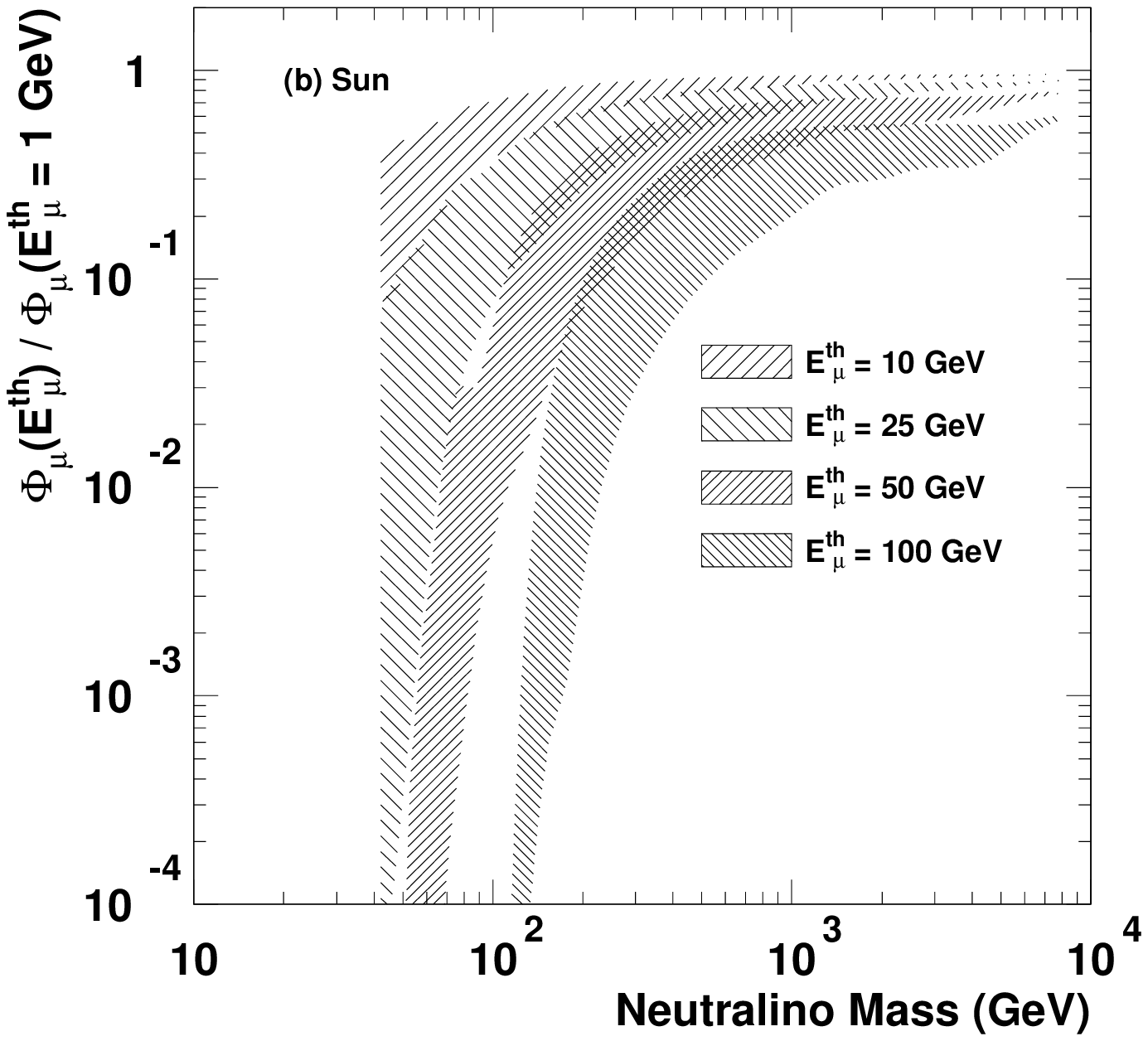,width=0.49\textwidth}}
\caption{The ratio of the muon fluxes for different thresholds to those
with a threshold of 1 GeV\@.}
\label{fig:reasurmx}
\end{figure}

In Fig.~\ref{fig:reasurmx} we show the ratio of the muon flux with 
different thresholds $E_{\mu}^{th}$ to those with a threshold of 1 
GeV\@. The width of the bands reflects the different degrees of softness 
of the neutrino spectra for a given neutralino mass.  The softer the 
neutrino spectrum, the more we lose by increasing the threshold.  

We see that if the neutralino mass is above the threshold energy
by a fair amount, not too many events are lost by increasing the
energy threshold.  This is because the detection rate is determined by
the second moment of the neutrino energy (one power of energy because
of the rise of the neutrino cross section, and one power because of
the increasing muon path length). So the most energetic muons dominate
the rate. A higher threshold may even be more advantageous than a low
one, because it also reduces the background from atmospheric
neutrinos.

For the Sun, we see that there is always a loss even at the
highest masses.  This is due to the absorption of neutrinos in the
interior of the Sun, which softens the neutrino spectra.  When the
threshold exceeds 100 GeV, at least half of the signal from the Sun is
lost whatever the neutralino mass is.

As an example, if the neutralino mass were 200 GeV, increasing the
threshold from 1 to 100 GeV could decrease the signal by a factor of
between 10 and 1000.  On the other hand, if the threshold can be kept
at 25 GeV or below, we see that on the average only a factor of 2--3
is lost for a 200 GeV neutralino, from either the Sun or the Earth.
It is thus highly desirable to keep the threshold as low as possible
to keep the signal high.

The above discussion is true for muons traversing a thin detector.  
However, for $\cal O($1~km$^{3})$ neutrino telescopes, we would expect 
the event rates for contained events (i.e.\ tracks starting or 
stopping inside the detector) to be high also for masses below a few 
hundred GeV\@.  This can be expressed by an effective area that 
increases for low-mass neutralinos.  If this would be taken into 
account, the limits shown in Fig.~\ref{fig:r25mx} would go down by up 
to a factor of 10 at low masses.  We also note that for fully 
contained events (i.e.\ events both starting and stopping in the 
detector) it would be possible to get an energy estimate from the 
track length and in this case about another factor of 2 can be gained 
in sensitivity.

\section{Comparison with other signals} \label{sec:othersignals}

The uncertainty in the capture rates governing the muon flux enter in
a similar way in the calculation of the rates of direct detection. In
Fig.~\ref{fig:r25mx} we show with different symbols models that can be
probed with current direct detection experiments and what could be
obtained with a factor of 10 increase in sensitivity. In the Earth,
the correlation is fairly strong, whereas it is weaker in the Sun. The
reason is that the capture rate in the Sun depends on the
spin-dependent scattering cross section as well as the
spin-independent one.

We can also look for neutralinos by searching for their annihilation
products from annihilation in the halo. The most interesting sources
are gamma lines, continuum gammas, antiprotons and
positrons. The correlation with these signals is fairly weak and they
thus represent fairly complementary methods of searching for
neutralino dark matter. We do get higher uncertainties from the halo
profile and propagation (for charged particles) though.

One of the earliest precursors of the MSSM may be the discovery of the
Higgs boson at accelerators, where the lightest neutral Higgs scalar
$H_{2}^0$ in supersymmetric models hardly can be heavier than 130 GeV
after loop corrections (Carena et al., 1995) have been included.  We
find models with high muon rates (regardless of the threshold) all the
way up to the heaviest $H_{2}^0$ allowed in the MSSM.  An MSSM Higgs
boson of mass near the 130 GeV limit will not be detectable by LEP II,
and may require several years of LHC running for its discovery.

\section{Conclusions}

We have seen that the higher threshold of the new generation of
neutrino telescopes reduces the rates for low-mass neutralinos whereas
the suppression is less severe for high-mass models.
For muons from the Earth, the suppression means that neutrino 
telescopes will have some difficulties to compete with direct 
detection methods.  For the Sun the situation is different as the 
spin-dependent cross section has a larger spread, and there do not yet 
exist direct detectors of large sensitivity.  From the point of view 
of neutralino search, the optimum design of a neutrino telescope would 
have a low muon energy threshold and a good sensitivity to search for 
a signal from the direction of the Sun.

Various methods of detecting supersymmetric 
dark matter probe complementary regions of parameter space, and are 
therefore all worth pursuing experimentally.  The dark matter problem 
remains one of the outstanding problems of basic science.  Maybe the 
first clues to its solution will come from the large new neutrino 
telescopes presently being planned.

\section*{Acknowledgements}

LB was supported by the Swedish Natural Science Research 
Council (NFR).  We thank Piero Ullio for discussions.  This work was 
supported with computing resources by the Swedish Council for High 
Performance Computing (HPDR) and Parallelldatorcentrum (PDC), Royal 
Institute of Technology.


\vspace{1ex}
\begin{center}
{\Large\bf References}
\end{center}
%
\setlength{\baselineskip}{0.9\baselineskip}
Alam, M.S.\ et al., \prl{74}{1995}{2885}.\\
Ammar, R.\ et al.\, \prl{71}{1993}{674}.\\
Bergstr{\"o}m, L., Edsj{\"o}, J.\ \& Gondolo, P.\ 1998, Phys.\ Rev.\ 
{\bf D58}, 103519\\
Bergstr\"om, L., Edsj\"o, J.\ \& Kamionkowski, M.,
\app{7}{1997}{147}.\\
Bernabei, R.\ et al.\ ({\sc Dama} Collaboration), 
\plb{389}{1996}{757}.\\
Carena, M.\ et al., 
\plb{355}{1995}{209}.\\
Carr, J., 1998, talk given March 31, 1998,
\href{http://alephwww.cern.ch/ALPUB/seminar/carrlepc98/index.html}
{http://alephwww.cern.ch/ALPUB/seminar/carrlepc98/index.html},
Preprint ALEPH 98-029, 1998 winter conferences,
\href{http://alephwww.cern.ch/ALPUB/oldconf/oldconf.html}
{http://alephwww.cern.ch/ALPUB/oldconf/oldconf.html}.\\
Halzen, F., 1999, these proceedings.\\
Honda, M.\ et al., \prd{52}{1995}{4985}.\\
Ingelman, G.\ \& Thunman, M., \prd{54}{1996}{4385}.\\
Jungman, G, Kamionkowski, M.\ \& Griest, K.\ 1996, Phys.\ Rep.\ 
{\bf 267} 195.\\
Edsj{\"o}, J., 1997, PhD Thesis, Uppsala University, hep-ph/9704384\\
Edsj{\"o}, J.\ \& Gondolo, P., \prd{56}{1997}{1879}.\\
Seckel, D., Stanev, T.\ \& Gaisser, T.K., \apj{382}{1991}{651}\\
Sj\"{o}strand, T., \cpc{82}{1994}{74}.\\
~\\[-2ex]
\centerline{\it For a more detailed list of references, see
Bergstr{\"o}m, L., Edsj{\"o}, J.\ \& Gondolo, P., 1998}

\end{document}